\documentstyle[12pt,a4,epsf]{article}

\makeatletter
\@addtoreset{equation}{section}
\makeatother

\newcommand{\VEV}[1]{\left\langle #1 \right\rangle}

\newcommand{\nn}{\nonumber}

\newcommand{\MZ}{M_{\mrm Z}}

\newcommand{\order}[1]{${\cal O}(#1)$}
\newcommand{\GeV}{\mbox{GeV}}

\newcommand{\alphas}{\alpha_s(\MZ)}

\newcommand{\Gsm}{$SU(3)_C\times SU(2)_L\times U(1)_Y$}
\newcommand{\Ga}{$SU(3)_C\times SU(2)_L\times SU(2)_R
                 \times U(1)_{B-L}$}
\newcommand{\E}[1]{$E_#1$}

\newcommand{\cc}[1]{\overline{#1}}
\newcommand{\sub}[1]{$_{\mrm{#1}}$}
\newcommand{\eff}{{\rm {eff}}}

\newcommand{\bequ}{\begin{equation}}
\newcommand{\eequ}{\end{equation}}
\newcommand{\beqn}{\begin{eqnarray}}
\newcommand{\eeqn}{\end{eqnarray}}
\newcommand{\bctr}{\begin{center}}
\newcommand{\ectr}{\end{center}}
\newcommand{\Ls}{\left(}
\newcommand{\Rs}{\right)}

\newcommand{\vsp}[1]{\vspace {#1cm}}
\newcommand{\half}{{1\over2}}
\newcommand{\mrm}{\rm}
\newcommand{\II}{I$\!$I}
\newcommand{\III}{I$\!$I$\!$I}

\begin{document}
\begin{titlepage}

\begin{flushright}
hep-ph/0205185\\
KUNS-1786\\
\today
\end{flushright}

\vspace{4ex}

\begin{center}
{\large \bf
Two-Loop Analysis of Gauge Coupling Unification \\
with Anomalous $U(1)$ Symmetry
and Proton Decay
}

\vspace{6ex}

\renewcommand{\thefootnote}{\alph{footnote}}
Nobuhiro {\sc Maekawa}\footnote
{E-mail: maekawa@gauge.scphys.kyoto-u.ac.jp
}
and 
Toshifumi {\sc Yamashita}\footnote{
E-mail: yamasita@gauge.scphys.kyoto-u.ac.jp
}

\vspace{4ex}
{\it Department of Physics, Kyoto University, Kyoto 606-8502, Japan}\\
\end{center}

\renewcommand{\thefootnote}{\arabic{footnote}}
\setcounter{footnote}{0}
\vspace{6ex}

\begin{abstract}
Recently, a new mechanism, which explains why the three gauge 
coupling constants meet at a certain scale in the minimal 
supersymmetric standard model, has been proposed in a scenario of 
grand unified theories with anomalous $U(1)_A$ gauge symmetry. 
It is a non-trivial result that although there are many superheavy 
fields whose mass scales are below the unification scale, 
this mechanism explains this fact by means of one-loop 
renormalization group equations.
Since the unification scale generically becomes below the usual 
GUT scale, $2\times 10^{16}$ GeV, and proton decay via dimension 5 
operators is suppressed, the scenario predicts that proton decay via 
dimension 6 operators, $p\rightarrow e \pi^0$, will be observed in 
the near future. In this paper, we attempt to estimate a reasonable 
range of values of the lifetime of the proton predicted within this 
scenario by using a two-loop renormalization group calculation and 
the ambiguities of \order1 coefficients.



\end{abstract}

\end{titlepage}

\section{Introduction}
Low energy supersymmetry (SUSY), which was originally introduced for 
stabilization of the weak scale, plays a critical role in explaining
the hierarchical gauge couplings of the standard model
in the context of $SU(5)$ grand unified theory (GUT). 
It is a non-trivial
result that the three gauge couplings meet at a GUT scale 
$\Lambda_G\sim 2\times 10^{16}$ GeV if a reasonable
SUSY breaking scale is assumed. 
However, the experimental lower bound of the proton 
lifetime via dimension 5 operators has become so severe that many GUTs
that naturally realize coupling unification, for example, the
minimal $SU(5)$ GUT, have been rejected.
\cite{Murayama,Goto,lattice,hisano}%
\footnote{
Recently, it was pointed out that allowing arbitrary soft masses and 
fermion and sfermion mixing, there is still only a small portion of 
parameter space consistent with experimental results 
in the `decoupling' region.
\cite{Bajc}
}
This is a kind of puzzle 
in the SUSY 
GUT scenario. Of course, if we consider the coincidence of the three 
gauge couplings at some scale to be accidental, then this puzzle is 
not difficult to solve. For example, after suppression of 
dimension 5 proton decay, the gauge couplings can be caused to meet 
at a certain scale by tuning the mass scales under the  GUT scale. 
However, if  the coincidence of these couplings is not accidental, 
solving this puzzle is not so simple, although some solutions have 
been proposed. 
Most of these solutions aim to show that the minimal SUSY 
standard model (MSSM) is realized under the GUT scale $\Lambda_G$
with the dimension 5 proton decay suppressed or forbidden. 
There are several solutions of this type in the context of 
extra dimensions. 
One employs parity assignment,
which forbids dimension 5 operators.\cite{kawamura,hall}
The other employs wave function 
suppression due to localization of quark and lepton fields.
\cite{kakizaki,maru}
Even in the 
context of four-dimensional field theory, solving this puzzle is 
possible by introducing a special vacuum structure of one or two adjoint 
Higgs fields of $SO(10)$.
\cite{babu}
However, recently, another type of solution has been proposed
\cite{maekawa3} in the context
of GUT with anomalous $U(1)$ gauge symmetry,\cite{U(1)} whose anomaly 
is cancelled by the Green-Schwarz mechanism.\cite{GS} It is surprising 
that in the GUT scenario with a simple unification group, 
gauge coupling unification is realized generically, 
despite the fact that many superheavy 
fields, whose mass spectrum does not respect the GUT symmetry, become 
lighter than the unification scale $\Lambda_A$ and there are 
several gauge symmetry breaking scales. 
This is because the mass spectrum of superheavy fields and the symmetry
breaking scales are determined by the anomalous $U(1)_A$ charges, and
most of the charges are cancelled under the conditions of 
coupling unification.
(The unique exception is the charge of the doublet Higgs.)
Moreover, the GUT scenario has many interesting features which we now 
describe.
\cite{maekawa3,maekawa,maekawa2,BM,MY}
1) The interaction is generic in the sense that all the interactions 
that are allowed by the symmetry are introduced. 
Therefore, once we fix the field
content with their quantum numbers (integers), all the interactions are
determined, except the coefficients of order 1.
2) It naturally solves the so-called 
doublet-triplet (DT) splitting problem, \cite{DTsplitting} 
using the Dimopoulos-Wilczek (DW) mechanism.
\cite{DW,BarrRaby}  
3) It reproduces the  realistic structure of the quark and lepton 
mass matrices, including neutrino bi-large mixing,\cite{SK}
 using the 
Froggatt-Nielsen (FN) mechanism.\cite{FN}  
4) The anomalous $U(1)_A$ accounts for the hierarchical structure of 
the symmetry breaking scales  and the masses of 
heavy particles. 
5) All the fields, except those of the minimal 
SUSY standard model (MSSM), can become heavy.
6) The gauge couplings are unified just below the usual GUT scale, 
$\Lambda_G\sim 2\times 10^{16}$ GeV.
7) In spite of the lower unification scale, proton decay via 
dimension 6 operators, $p\rightarrow e^+\pi^0$, is still within 
the experimental bound and therefore we expect to observe 
proton decay in the near future.
8) The cutoff scale is lower than the Planck scale.
9) The $\mu$ problem is also solved.

In the above-mentioned scenario, one of the most interesting 
predictions regards proton decay. 
Because the dimension 5 operators are suppressed, the main decay mode
of proton decay is due to dimension 6 operators. Therefore, a more 
accurate estimate of the unification scale or the cutoff scale is 
important in order to obtain a more accurate prediction of the 
lifetime of the proton. 
In this paper, we determine an allowed range of values of the 
cutoff scale for several GUT models by means of two-loop 
renormalization group equations (RGEs) and using 
the freedom of \order1 coefficients.

\section{Gauge coupling unification (one-loop renormalization group)}
\label{review}
In our scenario, because generic interactions are introduced, 
the order of every coefficient is determined by anomalous $U(1)_A$ 
charges. Therefore, the gauge symmetry breaking scales and the mass 
spectrum of superheavy fields are also determined by these charges. 
Thus we can examine whether the gauge couplings meet at the GUT scale 
once we determine all the charges.
This is a consistency check of our scenario, and it has been 
shown that it is realized in a non-trivial way. In this section, we 
review this point using one-loop RGEs.

First, we note that the symmetry breaking scales are determined
by anomalous $U(1)_A$ charges. Generically, 
the vacuum expectation value (VEV)
of the gauge singlet operator $O$ is determined by its charge $o$ as
\begin{equation}
\VEV{O}\sim \lambda^{-o}.
\end{equation}
(Here $\lambda$ is the ratio of the cutoff scale $\Lambda$ to the VEV of
the Froggatt-Nielsen field $\Theta$. In our scenario, we adopt
$\lambda\sim 0.22$. )
Actually, in our scenario, the VEV of the adjoint field $A$ of
$SO(10)$ becomes of the Dimopoulos-Wilczek type :
\begin{equation}
\VEV{A}=i\tau_2\times {\rm diag}
(v,v,v,0,0).
\end{equation}
This fact plays an important role in realizing doublet-triplet splitting,
and the scale is determined by the charges $a$ as 
$v\sim \lambda^{-a}\Lambda$. 
Throughout this paper we  denote all 
superfields by uppercase letters and their anomalous $U(1)_A$ charges
by the corresponding lowercase letters. 
We often use units in which $\Lambda=1$.
The VEV of $A$ breaks $SO(10)$ into 
$SU(3)_C\times SU(2)_L\times SU(2)_R\times U(1)_{B-L}$.
The VEVs of the spinors $C({\bf 16})$ and 
$\bar C({\bf\overline{16}})$, 
which break $SU(2)_R\times U(1)_{B-L}$ into $U(1)_Y$, are
obtained from $\VEV{\bar CC}\sim \lambda^{-(c+\bar c)}$ and 
the $D$-flatness condition $|\VEV{C}|=|\VEV{\bar C}|$ as
\begin{equation}
|\VEV{C}|=|\VEV{\bar C}|\sim \lambda^{-(c+\bar c)/2}.
\end{equation}
Because $|\VEV{C}|<|\VEV{A}|$ is required to realize doublet-triplet
splitting,
at the scale 
$\Lambda_A\equiv\VEV{A}\sim \lambda^{-a}$, 
the $SO(10)$ gauge group is broken into
$SU(3)_C\times SU(2)_L\times SU(2)_R\times U(1)_{B-L}$, which is broken
into the standard gauge group $SU(3)_C\times SU(2)_L\times U(1)_Y$ 
at the scale $\Lambda_C\equiv\VEV{C}\sim \lambda^{-(c+\bar c)/2}$. 

Second, we explain how the mass spectrum of superheavy fields is 
determined by anomalous $U(1)_A$ charges. 
Using the definitions of the fields $Q({\bf 3,2})_{\frac{1}{6}}$,
$U^c({\bf \bar 3,1})_{-\frac{2}{3}}$, $D^c({\bf \bar 3,1})_{\frac{1}{3}}$,
$L({\bf 1,2})_{-\frac{1}{2}}$, $E^c({\bf 1,1})_1,N^c({\bf 1,1})_0$ 
and $X({\bf 3,2})_{-\frac{5}{6}}$, along with their conjugate fields, and
$G({\bf 8,1})_0$ and $W({\bf 1,3})_0$ with the standard gauge symmetry,
under $SO(10)\supset SU(5) \supset SU(3)_C\times SU(2)_L\times U(1)_Y$,
the spinor ${\bf 16}$, vector ${\bf 10}$ and adjoint ${\bf 45}$
of $SO(10)$
are decomposed as
\begin{eqnarray}
{\bf 16}&\rightarrow &
\underbrace{[Q+U^c+E^c]}_{\bf 10}+\underbrace{[D^c+L]}_{\bf \bar 5}
+\underbrace{N^c}_{\bf 1},\\
{\bf 10}&\rightarrow &
\underbrace{[D^c+L]}_{\bf \bar 5}+\underbrace{[\bar D^c+\bar L]}_{\bf 5},\\
{\bf 45}&\rightarrow &
\underbrace{[G+W+X+\bar X+N^c]}_{\bf 24}
+\underbrace{[Q+U^c+E^c]}_{\bf 10}
+\underbrace{[\bar Q+\bar U^c+\bar E^c]}_{\bf \overline{10}}
+\underbrace{N^c}_{\bf 1}.
\end{eqnarray}
A straightforward calculation of the mass matrices $\bar M_I$ of 
the superheavy fields $I=Q$, $U^c$, $E^c$, $D^c$, $L$, $G$, $W$ and 
$X$ shows that
\begin{equation}
\det \bar M_I = \lambda^{\mbox{$\sum_i c_i$}},
\end{equation}
where the quantities $c_i$ are the anomalous $U(1)_A$ charges of 
the superheavy fields.

Third, we carry out an analysis based on the RGEs up to one loop. 
The conditions of gauge coupling unification are 
\begin{equation}
\alpha_3(\Lambda_A)=\alpha_2(\Lambda_A)=
\frac{5}{3}\alpha_Y(\Lambda_A)\equiv\alpha_1(\Lambda_A),
\end{equation}
where 
$\alpha_1^{-1}(\mu>\Lambda_C)\equiv 
\frac{3}{5}\alpha_R^{-1}(\mu>\Lambda_C)
+\frac{2}{5}\alpha_{B-L}^{-1}(\mu>\Lambda_C)$.
Here $\alpha_X=\frac{g_X^2}{4\pi}$, and 
the parameters $g_X\, (X=3,2,R,B-L,Y)$ are the gauge couplings of 
$SU(3)_C$, $SU(2)_L$, $SU(2)_R$, $U(1)_{B-L}$ and $U(1)_Y$, 
respectively.

The gauge couplings at the scale $\Lambda_A$ are roughly given by
\begin{eqnarray}
\alpha_1^{-1}(\Lambda_A)&=&\alpha_1^{-1}(M_{SB})
+\frac{1}{2\pi}\left(b_1\ln \left(\frac{M_{SB}}{\Lambda_A}\right)
+\sum_i \Delta b_{1i}\ln \left(\frac{m_i}{\Lambda_A}\right)
-\frac{12}{5}\ln \left(\frac{\Lambda_C}{\Lambda_A}\right)\right), 
\label{alpha1} \nn \\
&&  \\
\alpha_2^{-1}(\Lambda_A)&=&\alpha_2^{-1}(M_{SB})
+\frac{1}{2\pi}\left(b_2\ln \left(\frac{M_{SB}}{\Lambda_A}\right)
+\sum_i \Delta b_{2i}\ln \left(\frac{m_i}{\Lambda_A}\right)\right), \\
\alpha_3^{-1}(\Lambda_A)&=&\alpha_3^{-1}(M_{SB})
+\frac{1}{2\pi}\left(b_3\ln \left(\frac{M_{SB}}{\Lambda_A}\right)
+\sum_i \Delta b_{3i}\ln \left(\frac{m_i}{\Lambda_A}\right)\right), 
\end{eqnarray}
where $M_{SB}$ is a SUSY breaking scale, 
$(b_1,b_2,b_3)=(33/5,1,-3)$ are the 
renormalization group coefficients
for the minimal SUSY standard model (MSSM),
and $\Delta b_{ai}\ (a=1,2,3)$ are the corrections to the coefficients 
from the massive fields with mass $m_i$.
The last term in Eq.~(\ref{alpha1}) is from the breaking 
$SU(2)_R\times U(1)_{B-L}\rightarrow U(1)_Y$ caused by the VEV $\VEV{C}$.
Because the gauge couplings at the SUSY breaking scale $M_{SB}$
are given by
\bequ
\alpha_i^{-1}(M_{SB})=\alpha_G^{-1}(\Lambda_G)
+\frac{1}{2\pi}\left(b_i\ln\left(\frac{\Lambda_G}{M_{SB}}
               \right)\right)\quad,\quad(i=1, 2, 3)
\eequ
where $\alpha_G^{-1}(\Lambda_G)\sim 25$ and 
$\Lambda_G\sim 2\times 10^{16}$ GeV, the above conditions for unification
can be rewritten as
\begin{eqnarray}
&&b_1\ln \left(\frac{\Lambda_A}{\Lambda_G}\right)
+\Sigma_I\Delta b_{1I}\ln \left(\frac{\Lambda_A^{\bar r_I}}{\det \bar M_I}
\right)
-\frac{12}{5}\ln \left(\frac{\Lambda_A}{\Lambda_C}\right) \\
&=&b_2\ln \left(\frac{\Lambda_A}{\Lambda_G}\right)
+\Sigma_I\Delta b_{2I}\ln\left(\frac{\Lambda_A^{\bar r_I}}{\det \bar M_I}
\right) \\
&=&b_3\ln \left(\frac{\Lambda_A}{\Lambda_G}\right)
+\Sigma_I\Delta b_{3I}\ln\left(\frac{\Lambda_A^{\bar r_I}}{\det \bar M_I}
\right).
\end{eqnarray}
Here, 
$\bar r_I$ represents the ranks of the mass matrices of 
the superheavy fields, $\bar M_I$. 
The corrections to the renormalization coefficients $\Delta b_{aI}$ are
given in the following table:
\begin{center}
\begin{tabular}{|c|c|c|c|c|c|c|c|c|}
\hline
$I$      & $Q+\bar Q$ & $U^c+\bar U^c$ & $E^c+\bar E^c$ & $D^c+\bar D^c$ 
         & $L+\bar L$ & $G $& $W$ & $X+\bar X$ \\
\hline
$\Delta b_{1I}$ & $\frac{1}{5}$& $\frac{8}{5}$& $\frac{6}{5}$& $\frac{2}{5}$
         & $\frac{3}{5}$ & 0 &  0 & 5 \\
\hline
$\Delta b_{2I}$ & 3 & 0 & 0 & 0 & 1 & 0 & 2 & 3 \\
\hline
$\Delta b_{3I}$ & 2 & 1 & 0 & 1 & 0 & 3 & 0 & 2 \\ 
\hline
\end{tabular}
\end{center}
The unification conditions $\alpha_1(\Lambda_A)=\alpha_2(\Lambda_A)$,
$\alpha_1(\Lambda_A)=\alpha_3(\Lambda_A)$ and
$\alpha_2(\Lambda_A)=\alpha_3(\Lambda_A)$ can be rewritten as
\begin{eqnarray}
&&\left(\frac{\Lambda_A}{\Lambda_G}\right)^{14}
\left(\frac{\Lambda_C}{\Lambda_A}\right)^6
\left(\frac{\det \bar M_L}{\det \bar M_{D^c}}\right)
\left(\frac{\det \bar M_Q}{\det \bar M_{U}}\right)^4
\left(\frac{\det \bar M_Q}{\det \bar M_{E^c}}\right)^3
\left(\frac{\det \bar M_W}{\det \bar M_{X}}\right)^5  \nn \\ 
&&\qquad 
=\Lambda_A^{-\bar r_{D^c}+\bar r_L-4\bar r_{U^c}-3\bar r_{E^c}+7\bar r_Q
-5\bar r_X+5\bar r_W}, \\
&&\left(\frac{\Lambda_A}{\Lambda_G}\right)^{16}
\left(\frac{\Lambda_C}{\Lambda_A}\right)^4
\left(\frac{\det \bar M_{D^c}}{\det \bar M_{L}}\right)
\left(\frac{\det \bar M_Q}{\det \bar M_{U}}\right)
\left(\frac{\det \bar M_Q}{\det \bar M_{E^c}}\right)^2
\left(\frac{\det \bar M_G}{\det \bar M_{X}}\right)^5  \nn \\ 
&&\qquad =
\Lambda_A^{-\bar r_{L}+\bar r_{D^c}-\bar r_{U^c}-2\bar r_{E^c}+3\bar r_Q
-5\bar r_X+5\bar r_G}, \\
&&\left(\frac{\Lambda_A}{\Lambda_G}\right)^{4}
\left(\frac{\det \bar M_{D^c}}{\det \bar M_{L}}\right)
\left(\frac{\det \bar M_U}{\det \bar M_{Q}}\right)
\left(\frac{\det \bar M_G}{\det \bar M_{W}}\right)^2
\left(\frac{\det \bar M_G}{\det \bar M_{X}}\right) \nn \\ 
&&\qquad =\Lambda_A^{-\bar r_{L}+\bar r_{D^c}-\bar r_{Q}+\bar r_{U}-2\bar r_W
-\bar r_X+3\bar r_G}.
\end{eqnarray}
Note that the above conditions are dependent only on the ratio 
of the determinants of the mass matrices that are included 
in the same multiplet of $SU(5)$ and on the symmetry breaking 
scales $\Lambda_A$ and $\Lambda_C$.
If all the component fields in a multiplet were superheavy, the
above ratios would be of order 1. However, because some of the component
fields, for example massless Higgs doublets or Nambu-Goldstone modes, 
do not appear 
in the mass matrices generically, the above ratios are dependent only 
on the charges of these massless modes. 
If all the fields other than those in MSSM become
superheavy, the above ratios are easily estimated as
\begin{eqnarray}
\frac{\det \bar M_L}{\det \bar M_{D^c}}&\sim & \lambda^{-(h_u+h_d)}, \\
\frac{\det \bar M_Q}{\det \bar M_{E^c}}\sim 
\frac{\det \bar M_{U^c}}{\det \bar M_{E^c}}&\sim &\lambda^{c+\bar c-2a}, \\
\frac{\det \bar M_G}{\det \bar M_X}\sim 
\frac{\det \bar M_W}{\det \bar M_X}&\sim &\lambda^{-2a},
\end{eqnarray}
where $h_u$ and $h_d$ are the anomalous $U(1)_A$ charges of 
the massless Higgs doublets $H_u$ and $H_d$, respectively.
Then the conditions for coupling unification become 
\begin{eqnarray}
\Lambda &\sim & \lambda^{\frac{h_u+h_d}{14}}\Lambda_G,\label{cond1}\\
\Lambda &\sim &\lambda^{-\frac{h_u+h_d}{16}}\Lambda_G,\label{cond2} \\
\Lambda &\sim &\lambda^{-\frac{h_u+h_d}{4}}\Lambda_G.\label{cond3}
\end{eqnarray}
Therefore the unification conditions become $h_u+h_d\sim 0$, and thus 
the cutoff scale must be taken as $\Lambda\sim \Lambda_G$.
It is obvious that if the cutoff scale were some other scale 
(for example, the Planck scale), then in MSSM the three gauge couplings 
would meet at that scale.
This implies that in this scenario it is not accidental that the 
three gauge couplings meet at some scale in MSSM.
Note that these results are independent of the details of 
the Higgs sector, because the requirement that all the fields other 
than those in MSSM become superheavy determines the field 
content of the massless fields, whose charges are important 
to examine whether the gauge couplings meet at the unification scale 
$\Lambda_A$. The above argument can also be applied to 
the scenario of $E_6$ unification, though instead
of the usual doublet Higgs charge $h$, we have
to use effective Higgs charges, 
\begin{equation}
h_{\eff}\equiv h+\frac{1}{4}(\phi-\bar \phi),
\end{equation}
where $E_6$ is broken into $SO(10)$ by the VEV 
$|\VEV{\Phi}|=|\VEV{\bar \Phi}|\sim \lambda^{-\frac{1}{2}(\phi+\bar\phi)}$.

Note that the condition $h\sim 0$ does not mean that $h=0$, 
because there is an ambiguity involving
order 1 coefficients, and we have used only one-loop RGEs. 
However, the above analysis
is useful in providing a rough picture of the behavior in which 
we are interested.

\section{Proton decay}
In order to understand how proton decay via dimension 5 operators is 
suppressed in our scenario, we now examine the mass matrix 
of triplet Higgs.
From the interaction
\begin{equation}
W=\lambda^{h+h'+a}HAH'+\lambda^{2h'}H'^2,
\end{equation}
the mass matrices of doublet and triplet Higgs are
\begin{equation}
({\bf 5}_H, {\bf 5}_{H^\prime})
\left(\begin{array}{cc} 0 & \lambda^{h+h^\prime +a}\VEV{A} \\
                     \lambda^{h+h^\prime +a}\VEV{A} & \lambda^{2h^\prime}
      \end{array}\right)
\left(\begin{array}{c} {\bf \bar 5}_H \\ {\bf \bar 5}_{H^\prime}
\end{array}\right).
\end{equation}
Here $H$ and $H'$ represent {\bf 10} of $SO(10)$ and their charges
$h<0$ and $h'>0$. Note that the $H^2$ term is forbidden by the SUSY zero 
mechanism.
The colored Higgs obtain masses of order 
$\lambda^{h+h^\prime+a}\VEV{A}\sim \lambda^{h+h^\prime}$.
Because in general, $\lambda^{h+h^\prime}>\lambda^{2h^\prime}$,
proton decay is naturally suppressed. The effective colored
Higgs mass is estimated as
$m_c^{\eff}\sim (\lambda^{h+h^\prime})^2/\lambda^{2h^\prime}=\lambda^{2h}$, 
which is larger than the cutoff scale, because
$h<0$. If the cutoff scale is the usual GUT scale, 
$\Lambda_G=2\times 10^{16}$ GeV, 
the condition $m_c^{\eff}>10^{18}$ GeV requires $h\leq -2$.

Strictly speaking, the condition for gauge coupling unification, 
$h=0$, is not satisfied. However, we emphasize that the ambiguity of 
the coefficients of order 1 can naturally allow for the recovery of 
coupling unification.
In order to change the unification condition, 
we have to consider the terms including the adjoint field $A$ whose
VEV $\VEV{A}$ breaks GUT symmetry. In addition to the mass term $m_XXX$, 
higher-dimensional operators $\frac{m}{\Lambda} XAX$ must be 
taken into account.
In the usual GUT scenario, such a correction is much smaller than the 
tree level mass term, because of the suppression factor 
$\frac{\VEV{A}}{\Lambda}$.
Therefore, even if $m\sim m_X<\VEV{A}$ is realized due to some symmetry,
it is not easy to change the unification condition without 
finetuning or some special assumptions, 
for example, forbidding the mass term 
by some symmetry.
However, the GUT with anomalous $U(1)_A$ symmetry naturally realizes 
$m_X\sim m \frac{\VEV{A}}{\Lambda}$, because $m_X\sim \lambda^{2x}$,
$m\sim \lambda^{2x+a}$ and $\VEV{A}\sim \lambda^{-a}$. Moreover,
in addition to these terms, $XA^nX$ also contributes to the mass of 
the $X$ field after developing the VEV $\VEV{A}$ if these terms are 
allowed by the symmetry.
In other words, almost none of the coefficients respect GUT symmetry.
Therefore, by adjusting such coefficients we can naturally change 
the unification condition.

Finally, we recall proton decay via dimension 6 operators 
in our scenario.
Because the cutoff scale $\Lambda$ is around the usual GUT scale, 
$\Lambda_G=2\times 10^{16}$ GeV, and the unification scale becomes 
$\lambda^{-a}$ ($a<0$), proton decay via dimension 6 operators 
may be seen in future experiments.
If we roughly estimate the lifetime of the proton
using the formula in Ref.~\cite{hisano} and a recent result provided
by a lattice calculation for the hadron matrix element parameter
$\alpha$,\cite{lattice} we obtain 
\begin{equation}
\tau_p(p\rightarrow e\pi^0)\sim 2.8\times 10^{33}
\left(\frac{\Lambda_A}{5\times 10^{15}\ {\rm GeV}}\right)^4
\left(\frac{0.015({\rm GeV})^3}{\alpha}\right)^2  {\rm years}.
\end{equation}
This value%
\footnote{
  Note that this value is independent of the gauge coupling $g_{10}$ 
at the unification scale.
  This is because the ratio ${g_{10}\over m_X^2}$ is independent of 
$g_{10}$, because the mass of massive gauge bosons $m_X$ is given by 
$m_X=\sqrt2g_{10}v$.
}
 is near the present experimental limit.\cite{SKproton} 

\section{Two-loop analysis}
  As mentioned above, in our scenario, the cutoff scale $\Lambda$ is 
deeply related to the $SO(10)$ breaking scale 
$\VEV A\sim \lambda^{-a}$, and also to the mass of 
the massive gauge bosons, which cause proton decay. 
  Therefore, to estimate the lifetime of the proton, we need to know 
how large the cutoff scale is. 
  In addition, to realize somewhat accurate predictions, 
a two-loop analysis is necessary, 
because in our scenario there are many rather 
light superheavy particles, and therefore the coupling constants become 
larger than those of MSSM. 
  In fact, in the \E6 scenario, they sometimes become too strong to
  rely on the perturbation analysis. 
  However, in the $SO(10)$ scenario, 
the two-loop effect is still not excessively large, and we can treat it 
as a small correction to the one-loop analysis. 

In this section,  for several models, we investigate how large the cutoff 
scale can be by 
using the two-loop RGEs and 
altering \order1 coefficients in the range of 
$y_{\mrm{max}}^{-1}\leq y\leq y_{\mrm{max}}$ 
as free parameters.
The detailed procedure of this analysis is explained in Appendix 
\ref{recipe}.

\subsection{Models}
  There are several models that may work well.
  Among them we examine some representative models in 
Tables \ref{so10model} and \ref{e6model}.
\begin{table}
\caption
{ Anomalous $U(1)_A$ charges of the non-trivial representational 
 Higgs fields 
 ($A({\bf 45},-)$, $A'({\bf 45},-)$,
 $C({\bf 16},+)$, $\bar C({\bf \overline{16}},+)$, 
 $C'({\bf 16},-)$, $\bar C'({\bf \overline{16}},-)$, 
 $H({\bf 10},+)$ and $H'({\bf 10},-)$)
  and of the matter fields ($\Psi_1({\bf 16},+)$, $\Psi_2({\bf 16},+)$, 
  $\Psi_3({\bf 16},+)$ and $T({\bf 10},+)$) 
 in some models of the $SO(10)$ scenario. Here, the sign $+$ or $-$ 
represents the $Z_2$ parity, and
 we assign odd R-parity for the matter fields.
 }
\bctr
\begin{tabular}{|c||c|c|c|c|c|c|c|c||c|c|c|c|}\hline
    &$A$&$A'$&$C$&$\bar C$&$C'$&$\bar C'$&$H$&$H'$
    &$\Psi_1$&$\Psi_2$&$\Psi_3$&$T$\\\hline\hline
  i  &$-1$&3&$-4$&$-1$&3&6&$-3$&4&9/2&7/2&3/2&5/2\\\hline
  ii &$-1$&3&$-3$&0&2&5&$-2$&3&4&3&1&2\\\hline
  iii&$-1$&3&$-4$&$-1$&3&6&$-4$&5&5&4&2&3\\\hline
  iv &$-1$&3&$-7/2$&1/2&3/2&11/2&$-3$&4&9/2&7/2&3/2&2\\\hline
  v  &$-1$&3&$-1$&$-2$&4&3&$-6$&7&6&5&3&6\\\hline
  vi &$-1/2$&3/2&$-4$&$-1$&5/2&11/2&$-3$&7/2&9/2&7/2&3/2&5/2\\\hline
  vii&$-1/2$&3/2&$-1$&$-2$&7/2&5/2&$-6$&13/2&6&5&3&6\\\hline
\end{tabular}
\ectr
\label{so10model}
\end{table}

\begin{table}
\caption
{Anomalous $U(1)_A$ charges of the non-trivial representational 
Higgs fields 
 ($A({\bf 78},-)$, $A'({\bf 78},-)$,
 $\Phi({\bf 27},+)$, $\bar \Phi({\bf \overline{27}},+)$,
 $C({\bf 27},+)$, $\bar C({\bf \overline{27}},+)$, 
 $C'({\bf 27},-)$ and $\bar C'({\bf \overline{27}},-)$)
  and of the matter fields ($\Psi_1({\bf 27},+)$, $\Psi_2({\bf 27},+)$ 
  and $\Psi_3({\bf 27},+)$) 
 in some models of the $E_6$ scenario. Here, the sign $+$ or $-$ 
represents the $Z_2$ parity, and
 we assign odd R-parity for the matter fields.
 The MSSM doublet Higgs are contained in $\Phi$.
}
\bctr
\begin{tabular}{|c||c|c|c|c|c|c|c|c||c|c|c|}\hline
    &$A$&$A'$&$\Phi$&$\bar\Phi$&$C$&$\bar C$&$C'$&$\bar C'$
    &$\Psi_1$&$\Psi_2$&$\Psi_3$\\\hline\hline
  I   &$-1/2$&5/2&$-3$&2&$-5$&$-1$&13/2&13/2&9/2&7/2&3/2\\\hline
  \II &$-1/2$&5/2&$-3$&1&$-4$&$-1$&13/2&11/2&9/2&7/2&3/2\\\hline
  \III&$-1$&4&$-3$&2&$-6$&$-2$&7&8&9/2&7/2&3/2\\\hline
\end{tabular}
\ectr
\label{e6model}
\end{table}

\subsubsection{$SO(10)$ scenario}
 There are four groups of models, characterized by different 
values of $h$: ii with $h=-2$; i,iv and vi with $h=-3$; 
iii with $h=-4$; v and vii with $h=-6$. 
A model with a larger charge of the Higgs doublet
 can realize gauge coupling unification more naturally.
Therefore, in the models v and vii it is not so easy to realize gauge 
coupling unification, though these models have the advantage that 
the FCNC constraint is weaker because $\psi_1=t$. 
Half integer charges can play the same role as $Z_2$ parity or 
R-parity. For example, model i does not require R-parity, 
model vii does not require $Z_2$ parity, and model vi requires 
only either of R-parity or $Z_2$ parity.
  Models vi and vii have a value $-1/2$ for the charge of $A$, 
and therefore, because 
the unification scale is given by $\Lambda_A\sim \lambda^{-a}$, 
these models predict a more stable proton than the other models.
Note that the model ii has the possibility that proton decay via 
dimension 5 operators dominates that via dimension 6 operators.
\subsubsection{\E6 scenario}
  The model \III\, is the unique consistent model with $a=-1$. 
  Models I and \II\, are characterized by $a=-1/2$ and the ``effective 
charge'' $h_\eff=-17/4$ and $-4$, respectively. 
  These models of the \E6 scenario require only either R-parity or 
$Z_2$ parity, because half integer charges play the same role 
as these parities.
Also, they automatically  meet the condition for suppressing FCNC. 

\subsection{Results}
  For each model, we calculate the maximal value of the cutoff scale 
  $\Lambda$\sub{max}, with which gauge coupling unification is 
realized, setting the parameters $y$\sub{max}, $\lambda$ and 
$\alphas$ as in Tables \ref{so10result} and \ref{e6result}.
\begin{table}
\caption
{  $\Lambda$\sub{max} in units of $10^{16}\GeV$ for some models of 
the $SO(10)$ scenario. 
  Basically, the parameters are set as $y$\sub{max}$=2$, 
$\lambda=0.22$ and $\alpha^{-1}_s(\MZ)=8.44$, but in each case, 
one of these values is different from these, as indicated 
in the first row.
  The notation ``---'' means that there is no solution for 
gauge coupling unification.
}
\bctr
\begin{tabular}{|c||c|c|c|c|c|c|c|c|}\hline
    &&$\!y$\sub{max}$=2^2\!$&$\!y$\sub{max}$=2^\half\!$
     &$\!\lambda=0.20\!$&$\!\lambda=0.25\!$
     &$\!\alpha^{-1}_s=8.30\!$&$\!\alpha^{-1}_s=8.58\!$
     &$\!$1 loop$\!$\\\hline\hline
  i  & 3.00 & 15.3 &  --- & 2.80 & 3.32 & 3.45 & 2.61 & 3.89 \\\hline
  ii & 4.17 & 18.3 & 1.88 & 3.97 & 4.43 & 4.85 & 3.63 & 5.47 \\\hline
  $\!$iii$\!$& 1.95 & 10.0 &  --- & 1.77 & 2.22 & 2.24 & 1.68 & 2.53 \\\hline
  iv & 2.69 & 13.3 &  --- & 2.48 & 2.97 & 3.09 & 2.34 & 3.89 \\\hline
  v  &  --- & 3.78 &  --- &  --- &  --- &  --- &  --- &  --- \\\hline
  vi & 2.80 & 14.6 &  --- & 2.58 & 3.09 & 3.21 & 2.41 & 3.89 \\\hline
  $\!$vii$\!$&  --- & 3.93 &  --- &  --- &  --- &  --- &  --- &  --- \\\hline
\end{tabular}
\ectr
\label{so10result}
\end{table}
  In the \E6 scenario, we additionally adopt the constraint that the gauge 
  coupling constants do not become too large. 

\begin{table}
\caption
{  $\Lambda$\sub{max} in units of $10^{16}\GeV$ for some models of 
the \E6 scenario. 
  Basically, the parameters are set as $y$\sub{max}$=2$, 
$\lambda=0.22$ and $\alpha^{-1}_s(\MZ)=8.44$, and 
$\alpha_i(\Lambda_A)<1$ is required, but in each case, 
one of these values is different from these, as indicated 
in the first row.
  The notation ``---'' means that there is no solution for 
gauge coupling unification with the conditions considered. 
Most cases with no solution result from 
the condition $\alpha_i(\Lambda_A)<1$, which is much different from 
$SO(10)$ cases..
}
\bctr
\begin{tabular}{|c||c|c|c|c|c|c|c|c|}\hline
    &&$\!y$\sub{max}$=2^2\!$&$\!y$\sub{max}$=2^\half\!$
     &$\!\lambda=0.20\!$&$\!\lambda=0.25\!$
     &$\!\alpha^{-1}_s=8.30\!$&$\!\alpha^{-1}_s=8.58\!$
     &$\!$1 loop$\!$\\\hline\hline
  I   &\ 9.4 & 79.0 & 2.91 & 4.57 & 19.3 & 10.2 & 8.66 & 11.0 \\\hline
  \II & 11.7 & 99.3 & 3.63 & 5.75 & 21.1 & 12.5 & 10.7 & 11.6 \\\hline
  \III&  --- & 16.4 &  --- &  --- & 6.10 &  --- &  --- & 6.68 \\\hline
\end{tabular}
\ectr
\label{e6result}
\end{table}

\subsubsection{$SO(10)$ scenario}
Roughly speaking, as $h$ increases by 1, $\Lambda$\sub{max} 
increase by a factor of approximately 1.5. On the other hand,
for a single value of $h$ 
(models i, iv and vi), the values of $\Lambda$\sub{max} may differ 
from each other by only about 10\%, and $\Lambda$\sub{max} does not
depend strongly on the parameters $\lambda$ and $\alpha_s$, as seen
in Table \ref{so10result}. 
If we set $y_{\rm max}=2$, 
$\Lambda$\sub{max} is not far from the naively expected value
$\Lambda\sim 2\times 10^{16}$ GeV. 
Therefore, the prediction of the proton lifetime
is at most a factor of $\sim 20$ larger than the naively estimated value
$3\times 10^{33}$ years for the models with $a=-1$, and $5\times 10^{34}$
for the models with $a=-1/2$. Note that this factor of $\sim20$ 
represents the largest value allowed by the ambiguity in the value 
of the \order1 coefficients. This is an unlikely situation, and 
therefore we believe that the actual value of this factor is smaller.

\begin{figure}[tb]
\bctr
\vsp{-1}
\begin{tabular}{cc}
\leavevmode
\epsfxsize=100mm
\put(290,50){{\large ${\log \Lambda (\GeV)}$}}
\put(5,243){{\Large ${\log_2 y_{\mrm max}}$}}
\put(197,217){{\normalsize \bf{vii}}}
\put(244,217){{\normalsize \bf{iii}}}
\put(268,217){{\normalsize \bf{i}}}
\put(277,217){{\normalsize \bf{ii}}}
\epsfbox{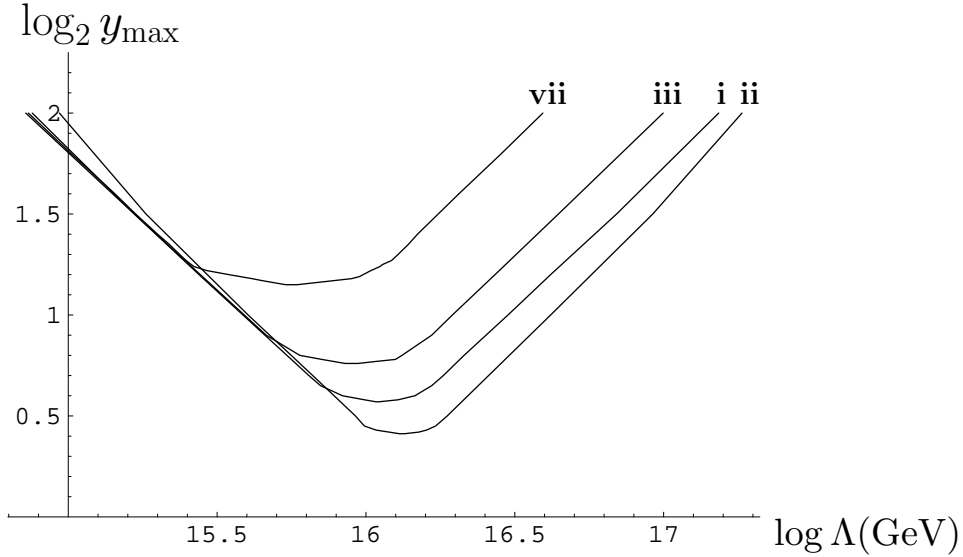}
\vspace{-2cm}
\end{tabular}
\ectr
\caption
{ $y$\sub{max} vs $\Lambda$\sub{min} and $\Lambda$\sub{max}. 
  Here, the four curves corresponding to the four models, 
i, ii, iii and vii
($h=-3, -2, -4, -6$, respectively), 
are plotted together. 
  The allowed region is the upper region of the curve
for each model. 
  Other parameters are set as $\lambda=0.22$ and 
$\alpha^{-1}_s(\MZ)=8.44$. 
}
\label{ymax}
\end{figure}

Unfortunately, the value $\Lambda$\sub{max} is strongly
dependent on $y_{\rm max}$, and unless we fix $y_{\rm max}$,
we cannot precisely predict the lifetime of the proton. 
For this reason, we analyse more
precisely the dependence on $y_{\rm max}$ in Fig.~\ref{ymax}.
It is found that the most probable $\Lambda$ becomes smaller
than the naively expected value $\Lambda_G\sim 2\times 10^{16}$
GeV, and this difference becomes larger as $h$ decreases. 
Furthermore, it is obvious that gauge coupling unification 
is more difficult to realize for smaller $h$. Therefore,
if we take account of the present limit on the proton lifetime 
provided by experiments, the models v and vii seem to be unrealistic.

  Finally, comparing the one- and two- loop results, it is seen that 
the two-loop effect in fact slightly worsens the situation, causing 
$\Lambda$\sub{\max} to decrease 
by a factor of $\sim1.4$, which corresponds to a prediction of 
the proton lifetime that is shorter by a factor of $\sim1/4$.

\subsubsection{\E6 scenario}
Since the $E_6$ scenario has a larger Higgs sector, the models in 
this case have more superheavy fields than in the $SO(10)$ scenario. 
This means that the $E_6$ 
models have larger degrees of freedom with regard to the values of 
the \order1 coefficients than do the $SO(10)$ models. 
Therefore, it is expected that gauge coupling unification will be
easier to realize in the \E6 models. 
Actually, as seen in Table \ref{e6result}, 
the maximal values of the cutoff $\Lambda_{\rm max}$ tend to
be larger here than in the $SO(10)$ cases. 
However, since the $E_6$ models have more superheavy fields, 
the gauge couplings around the unification scale tend to become 
so strong that the perturbative analysis is not reliable.
For this reason, we add an additional constraint at the unification 
scale $\Lambda_A$, 
\begin{equation}
\alpha_X(\Lambda_A)<1.
\end{equation}
Mainly because of this constraint, the results displayed 
in Table \ref{e6result} behave significantly differently.
In order to satisfy this condition that $\alpha_X$ be small, 
for many cases in the $E_6$ scenario, 
most of the degrees of freedom concerning the 
\order1 coefficients are lost. 
We can understand this from the $\lambda$ dependence elucidated 
in Table \ref{e6result}. 
Because the parameter $\lambda$ represents a unit of mass of the 
superheavy fields, $\alpha_X$ around the unification scale 
depends significantly on this parameter. Actually, as $\lambda$ 
becomes larger, the gauge couplings at the unification scale 
become smaller, and therefore 
more degrees of freedom of the \order1 coefficients remain unfixed 
and can be freely adjusted for the purpose of realizing 
gauge coupling unification. 
Through this effect, the increase of 
the maximal value of the cutoff $\Lambda_{\rm max}$ 
in the $E_6$ scenario becomes much larger than in the $SO(10)$ scenario.
In particular, in model \III, $\alpha_X$ becomes larger than 
in models I and \II, and hence more degrees of 
freedom are lost due to the requirement that $\alpha_X$ be small. 
Therefore it is more difficult to realize gauge coupling unification.
Actually, when $y_{\rm max}=2$, only the case with $\lambda=0.25$ 
can realize gauge coupling unification. 
If we tighten the constraint and require $\alpha_X$ to be smaller 
than unity at the cutoff scale $\Lambda$, even in models I and \II, 
coupling unification is difficult to realize.

\section{Discussion and summary}
  In the $SO(10)$ scenario, because the allowed range of values of 
the parameters $\lambda$ and  $\alphas$\  is limited, in this range, 
the maximal value of the cutoff $\Lambda_{\rm max}$ does not change 
greatly.
Because $\Lambda_{\rm max}$ is obtained by tuning the \order1 
coefficients to give the largest value, a reasonable value of 
$\Lambda$ would be less than the maximal value. 
If we fix $y_{\rm max}=2$, the maximal value of the cutoff does not 
differ by too much from the naively expected value,
$\Lambda\sim 2\times 10^{16}$ GeV, and therefore a realistic
prediction for proton decay can be considered to be 
not greatly different from the naive prediction, which is, 
for $a=-1$,
\bequ
\tau_p(p\rightarrow e\pi^0) \sim 2.8\times 10^{33}
\left(\frac{\Lambda_A}{5\times 10^{15}\ {\rm GeV}}\right)^4
\left(\frac{0.015({\rm GeV})^3}{\alpha}\right)^2  {\rm years},
\eequ
and for $a=-0.5$, 
\bequ
\tau_p(p\rightarrow e\pi^0)
\sim 4.5 \times 10^{34}
\left(\frac{\Lambda_A}{1\times 10^{16}\ {\rm GeV}}\right)^4
\left(\frac{0.015({\rm GeV})^3}{\alpha}\right)^2  {\rm years}.
\eequ
Unfortunately this prediction is strongly dependent on the unknown 
parameter for the \order1 coefficients, $y_{\rm max}$. 
If $y_{\rm max}=4$,
the upper bound of the prediction may be beyond the scope of future 
experiments. 

  In the \E6 scenario, on the other hand, because the coupling 
constants tend to be too large to allow a reliable perturbative 
analysis,  
we have added an artificial constraint to preserve the 
validity of our perturbative analysis.
Though the $E_6$ models have more degrees of freedom among the 
\order1 coefficients than do the $SO(10)$ models, most of them 
are lost to the requirement of suppressing the gauge couplings.
If we had some mechanism to suppress the gauge coupling constants
other than the degrees of freedom of the \order1 coefficients, then 
the $E_6$ models could much more easily 
realize gauge coupling unification. Actually, typical values of the 
maximal cutoff $\Lambda_{\rm max}$ in the \E6 cases are larger than 
in the $SO(10)$ cases.
It may be more natural in the $E_6$ scenario to assume that the gauge 
couplings are in the non-perturbative region. Though the perturbative 
analysis is not reliable in such cases, we may expect that the actual 
result is not so different from the perturbative prediction. 
This follows from the argument that if gauge coupling unification
in MSSM is not accidental, it may be natural to realize a 
situation similar to that in the perturbative region, 
which can explain gauge coupling unification in MSSM.

Note that there are other effects that can change the running of the 
gauge couplings that are not taken into account in the analysis 
presented in this paper. 
One effect is from SUSY breaking parameters. Generically, the 
SUSY breaking scale is not one scale, and therefore there are some 
effects contributing to the condition
for gauge coupling unification.
Here, we roughly estimate the contribution
in the case of model i of $SO(10)$ GUT.
If we set the masses of the colored superpartners to 1TeV and the 
masses of colorless superpartners to 100 GeV, then $\Lambda_{\max}$ 
decreases by a factor of $\sim$ 0.4. 
It would seem that, this is probably an overestimation.
For example, if we restrict the gaugino masses to satisfy 
the GUT relations
\begin{equation}
\frac{M_3}{\alpha_3}=\frac{M_2}{\alpha_2}=\frac{M_1}{\alpha_1},
\end{equation}
the factor by which $\Lambda_{\max}$ is decreased becomes 0.5-0.6. 
The contribution of the SUSY breaking parameters tends to decrease 
$\Lambda_{\max}$. Therefore, the estimations given in this paper 
still give conservative limits on the proton lifetime.
Another effect is from the lack of our knowledge about the values 
of the \order1 coefficients.
We have used 1 as the central value for the \order1 coefficients, but
this is only an assumption.
The result necessarily depends on this central value, 
especially in $E_6$ case, because the mass spectrum of superheavy 
fields depends on this value.
We have not used the ambiguities of the gauge symmetry breaking scale 
due to the \order1 coefficients. 
This effect can change not only the 
running of the gauge couplings directly 
but also the spectrum of superheavy 
fields if $\VEV{\bar CC}>\lambda^{-(c+\bar c)}$. From the 
non-renormalizable term $XX\bar CC$, by developing a non-vanishing 
VEV, the field $X$ can acquire a mass that is larger than expected. 
Though this effect exists only for fields lighter than 
$\lambda^{c+\bar c}$, it may suppress the gauge couplings at the
unification scale.

Finally, we emphasize that even if the upper bound for the 
prediction of the proton lifetime is beyond the scope of future 
experiments, a more likely prediction of our analysis is near 
the naive prediction, which is not very far from the present 
experimental bound.

\section*{Acknowledgements}
  T.Y. would like to thank Y.Okada for his encouragement.
  N.M. is supported in part by Grants-in-Aid for Scientific 
Research from the Ministry of Education, Culture, Sports, Science 
and Technology of Japan.
\appendix
\section{Recipe}\label{recipe}
  As mentioned in \S\ref{review}, in our scenario, the mass 
spectrum of every heavy Higgs is determined within the range 
allowed by the ambiguity of the \order1 coefficients. 
  Therefore we can easily calculate the second order $\beta$ 
function of the effective theories appropriate for each scale
(see Appendix \ref{beta}).
  If we use the $\cc{\mrm{DR}}$ scheme,\cite{DR} the naive step function 
approximation is good for connecting each gauge coupling constant 
of the neighboring effective theory, including the case in which the 
symmetries of these effective theories differ.\cite{step-fn}

  We adopt \Gsm\ for $\mu<\lambda^{-\frac{c+\bar c}{2}}\Lambda$, 
\Ga\ for 
$\lambda^{-\frac{c+\bar c}{2}}\Lambda<\mu<\lambda^{-a}\Lambda$, 
$SO(10)$ for $\lambda^{-a}\Lambda<\mu<\Lambda$ in the $SO(10)$ scenario and 
$\lambda^{-a}\Lambda<\mu<\lambda^{-\frac{\phi+\bar\phi}{2}}\Lambda$ 
in the \E6 scenario, and \E6 for
$\lambda^{-\frac{\phi+\bar\phi}{2}}\Lambda<\mu<\Lambda$ in the \E6 
scenario as the symmetries of the effective theories.  
  Here, we approximate the masses of massive gauge bosons 
as being the same as the symmetry switching scale. 

  Regarding the Yukawa coupling effect, we consider only that 
of the interaction directly related to the left- and right-handed 
top quark in terms of the relevant symmetry, {\it i.e.} $U^c_3Q_3H_u$
for \Gsm, ${Q_{\mrm R}}_3{Q_{\mrm L}}_3H_D$ for \Ga,
$\Psi_3({\bf{16}})\Psi_3({\bf{16}})H({\bf{10}})$ for $SO(10)$, and
$\Psi_3({\bf{27}})\Psi_3({\bf{27}})\Phi({\bf{27}})$ for \E6.
  For the scale dependence of the Yukawa coupling, we use one 
loop RGEs, which are sufficient for the purpose of 
investigating the flow of the gauge coupling constants. 

  We start from the central values of the $\cc{\mrm {MS}}$ gauge 
coupling constants at $\mu=\MZ$ given in Ref.\cite{pdg}, including 
$1\sigma$ ambiguity for $\alphas$.
  We set $M_t^{\cc{\mrm {MS}}}=165~\GeV, \tan\beta=5$, 
and $v=174~\GeV$, which correspond to $y_t=0.967$.
  From $\MZ$ to the SUSY breaking scale, we use the RGEs of the 
standard model, which contain three family 
fermions and one Higgs doublet.
  Since we do not specify the SUSY breaking mechanism, we fix
the SUSY breaking scale at 1 TeV and adopt a naive step function
approximation, except in the transformation from the $\cc{\mrm {MS}}$ 
scheme into 
$\cc{\mrm {DR}}$\cite{MS2DR}, as the SUSY breaking 
threshold effect.
  At the $SU(2)_R\times U(1)_{B-L}$ breaking threshold, we must 
transform \{$\alpha_Y, \alpha_2, \alpha_3$\} into \{$\alpha$\sub{B-L}, 
$\alpha$\sub R, $\alpha_2, \alpha_3$\}, in contrast to the one loop 
analysis.
  We set 
$\alpha^{-1}_{B-L}={3\over2}(\alpha^{-1}_Y - \alpha^{-1}_R)$ 
at $\mu=\lambda^{-(c+\bar c)/2}\Lambda$, 
and iteratively adjust $\alpha_R$ so that it is equal to $\alpha_2$ 
at the $SO(10)$ breaking scale $\lambda^{-a}\Lambda$.

  We investigate whether gauge coupling unification can be 
realized as explained above, by using the ambiguity of the 
\order1 coefficients, for some values of the parameters $\Lambda$, 
$y$\sub{max}, $\lambda$, and $\alphas$.
  In practice, we adjust  all the independent masses of the superheavy 
Higgs by factors of $y^{-1}_{\mrm max}$--$y$\sub{max}, instead of 
adjusting the \order1 coefficients.
  Then, if all the differences of the values of $\alpha_X^{-1}$ 
can be made smaller than $0.05$, we regard gauge coupling unification 
to be possible.

\section{Renormalization group equations}\label{beta}
  We use two-loop RGEs for the gauge coupling constants and one-loop 
RGEs for the top Yukawa coupling constant: 
\beqn
  \beta_i &=& {g_i^3\over16\pi^2}b_i 
            + {g_i^3g_j^2\over(16\pi^2)^2}b_{ij}
            + {g_i^3y_t^2\over(16\pi^2)^2}a_i,  \\
  \beta_t &=& {y_t^3\over16\pi^2}C
            + {y_tg_i^2\over16\pi^2}C_i.
\eeqn
  Here, $b_i, b_{ij}, a_i, C$ and $C_i$ are some constants.

  Considering only the top Yukawa interaction, as explained in Appendix %
\ref{recipe}, $a_i, C$ and $C_i$ are 
determined as follows.\cite{ano-dim} 
  For \Gsm~(non-SUSY)~($i=1, 2, 3$),
\bequ
  a_i = \Ls
  \begin{array}{c}
    17/10\\3/2\\2
  \end{array}
        \Rs, 
\quad
  C=9/2,
\quad
  C_i = \Ls
  \begin{array}{c}
    17/20\\9/4\\8
  \end{array}
        \Rs,
\eequ
  for \Gsm~($i=1, 2, 3$), 
\bequ
  a_i = \Ls
  \begin{array}{c}
    26/5\\6\\4
  \end{array}
        \Rs,
\quad
  C=6,
\quad
  C_i = \Ls
  \begin{array}{c}
    13/15\\3\\16/3
  \end{array}
        \Rs,  
\eequ
  for \Ga~($i=B-L, R, 2, 3$), 
\bequ
  a_i = \Ls
  \begin{array}{c}
    2\\12\\12\\8
  \end{array}
        \Rs,
\quad
  C=7,
\quad
  C_i = \Ls
  \begin{array}{c}
    1/3\\3\\3\\16/3
  \end{array}
        \Rs,  
\eequ
  for $SO(10)$ ($i=SO(10)$),  
\bequ
      a_i = 64, 
\quad C = 40,
\quad C_i = 63,
\eequ
and for \E6~($i=$\E6)
\bequ
      a_i = 180,  
\quad C = 60, 
\quad C_i = 104.
\eequ

 The constants $b_i$ and $b_{ij}$ are determined by the matter content of the 
effective theory.
  For example, for the standard model ($i=1, 2, 3$), 
which contains three families of 
fermions and one Higgs doublet,
\begin{eqnarray}
  b_i = \Ls
  \begin{array}{c}
    41/10\\-19/6\\-7
  \end{array}
        \Rs, 
&&
  b_{ij} = \Ls
  \begin{array}{ccc}
    199/50 & 27/10  & 44/5 \\
    9/10   & 35/6   & 12  \\
    11/10  & 9/2    & -26
  \end{array}
           \Rs ,
\end{eqnarray}
and for the MSSM ($i=1, 2, 3$), which contains three family 
chiral superfields and two Higgs doublet chiral superfields,
\beqn
  b_i = \Ls
  \begin{array}{c}
    33/5\\1\\-3
  \end{array}
        \Rs, 
&&
  b_{ij} = \Ls
  \begin{array}{ccc}
    199/25 & 27/5  & 88/5 \\
    9/5    & 25    & 24  \\
    11/5  & 9    & 14
  \end{array}
           \Rs .
\eeqn
  The correction due to each representational chiral superfield 
is as follows.\cite{b_ij} \\
  For \Gsm\ ($i=1, 2, 3$), 
\beqn
  \Delta b_i = \Ls
  \begin{array}{c}
    1/5\\3\\2
  \end{array}
        \Rs, 
\quad
  \Delta b_{ij} = \Ls
  \begin{array}{ccc}
    1/75 & 3/5 & 16/15 \\
    1/5  & 21  & 16    \\
    2/15 & 6   & 68/3
  \end{array}
           \Rs 
&&
\mbox{for}\ Q+\bar Q, 
\\
  \Delta b_i = \Ls
  \begin{array}{c}
    8/5\\0\\1
  \end{array}
        \Rs, 
\quad
  \Delta b_{ij} = \Ls
  \begin{array}{ccc}
    128/75 & 0 & 128/15 \\
    0      & 0 & 0      \\
    16/15  & 0 & 34/3
  \end{array}
           \Rs 
&&
\mbox{for}\ U^c+\bar U^c, 
\\
  \Delta b_i = \Ls
  \begin{array}{c}
    6/5\\0\\0
  \end{array}
        \Rs, 
\quad
  \Delta b_{ij} = \Ls
  \begin{array}{ccc}
    72/25 & 0 & 0 \\
    0     & 0 & 0 \\
    0     & 0 & 0 
  \end{array}
           \Rs 
&&
\mbox{for}\ E^c+\bar E^c, 
\\
  \Delta b_i = \Ls
  \begin{array}{c}
    2/5\\0\\1
  \end{array}
        \Rs, 
\quad
  \Delta b_{ij} = \Ls
  \begin{array}{ccc}
    8/75 & 0 & 32/15 \\
    0    & 0 & 0     \\
    4/15 & 0 & 34/3
  \end{array}
           \Rs 
&&
\mbox{for}\ D^c+\bar D^c, 
\\
  \Delta b_i = \Ls
  \begin{array}{c}
    3/5\\1\\0
  \end{array}
        \Rs, 
\quad
  \Delta b_{ij} = \Ls
  \begin{array}{ccc}
    9/25 & 9/5 & 0 \\
    3/5  & 7   & 0 \\
    0    & 0   & 0
  \end{array}
           \Rs 
&&
\mbox{for}\ L+\bar L, 
\\
  \Delta b_i = \Ls
  \begin{array}{c}
    0\\0\\3
  \end{array}
        \Rs, 
\quad
  \Delta b_{ij} = \Ls
  \begin{array}{ccc}
    0 & 0 & 0 \\
    0 & 0 & 0 \\
    0 & 0 & 54 
  \end{array}
           \Rs 
&&
\mbox{for}\ G, 
\\
  \Delta b_i = \Ls
  \begin{array}{c}
    0\\2\\0
  \end{array}
        \Rs, 
\quad
  \Delta b_{ij} = \Ls
  \begin{array}{ccc}
    0 & 0  & 0 \\
    0 & 24 & 0 \\
    0 & 0  & 0
  \end{array}
           \Rs 
&&
\mbox{for}\ W,
\\
  \Delta b_i = \Ls
  \begin{array}{c}
    5\\3\\2
  \end{array}
        \Rs, 
\quad
  \Delta b_{ij} = \Ls
  \begin{array}{ccc}
    25/3 & 15 & 80/3 \\
    5    & 21 & 16   \\
    10/3 & 6  & 68/3
  \end{array}
           \Rs 
&&
\mbox{for}\ X+\bar X. 
\eeqn
For \Ga, first we define the abbreviation 
of each representation ($SU(3)_C, SU(2)_L, SU(2)_R$) 
$_{U(1)_{B-L}}$ as 
$Q_{ L}({\bf 3,2,1})_{\frac{1}{3}}$, 
$Q_{ R}({\bf \bar3,1,2})_{-\frac{1}{3}}$, 
$L_{ L}({\bf 1,2,1})_{-1}$, 
$L_{ R}({\bf 1,1,2})_1$, 
$H_{ T}({\bf 3,1,1})_{-\frac{2}{3}}$, 
$U({\bf 3,1,1})_{\frac{4}{3}}$, 
$X_Q({\bf 3,2,2})_{-\frac{2}{3}}$ and their conjugates, and 
$H_D({\bf 1,2,2})_0$, $G({\bf 8,1,1})_0$, 
$W_{ L}({\bf 1,3,1})_0$, $W_{ R}({\bf 1,1,3})_0$ 
and $N({\bf 1,1,1})_0$. 
\footnote{
  Here we write $U(1)_{B-L}$ charges with the usual normalization, 
but in the context of $SO(10)$ symmetry, they are multiplied by 
$\sqrt{3/8}$. 
}
  The spinor ${\bf 16}$, vector ${\bf 10}$ and adjoint ${\bf 45}$
of $SO(10)$ are decomposed as
\begin{eqnarray}
  {\bf 16} & \rightarrow &
    Q_{ L} + Q_{ R} + L_{ L} + L_{ R}, \\
  {\bf 10} & \rightarrow &
    H_{ T} + \bar H_{ T} + H_{ D}, \\
  {\bf 45} & \rightarrow &
    G + W_{ L} + W_{ R} + N + X_Q +\bar X_Q + U +\bar U.
\end{eqnarray}
Then, the corrections due to these representational chiral 
superfields are ($i=B-L, R, 2, 3$)
\beqn
  \Delta b_i = \Ls
  \begin{array}{c}
    1/2\\0\\3\\2
  \end{array}
        \Rs, 
\quad
  \Delta b_{ij} = \Ls
  \begin{array}{cccc}
    1/12 & 0 & 3/2 & 8/3 \\
    0    & 0 & 0   & 0   \\
    1/2  & 0 & 21  & 16   \\
    1/3  & 0 & 6   & 68/3
  \end{array}
           \Rs 
&&
\mbox{for}\ Q_{ L}+\bar Q_{ L}, 
\\
  \Delta b_i = \Ls
  \begin{array}{c}
    1/2\\3\\0\\2
  \end{array}
        \Rs, 
\quad
  \Delta b_{ij} = \Ls
  \begin{array}{cccc}
    1/12 & 3/2 & 0 & 8/3 \\
    1/2  & 21  & 0 & 16   \\
    0    & 0   & 0 & 0   \\
    1/3  & 6   & 0 & 68/3
  \end{array}
           \Rs 
&&
\mbox{for}\ Q_{ R}+\bar Q_{ R}, 
\\
  \Delta b_i = \Ls
  \begin{array}{c}
    3/2\\0\\1\\0
  \end{array}
        \Rs, 
\quad
  \Delta b_{ij} = \Ls
  \begin{array}{cccc}
    9/4 & 0 & 9/2 & 0 \\
    0   & 0 & 0   & 0 \\
    3/2 & 0 & 7   & 0 \\
    0   & 0 & 0   & 0   
  \end{array}
           \Rs 
&&
\mbox{for}\ L_{ L}+\bar L_{ L}, 
\\
  \Delta b_i = \Ls
  \begin{array}{c}
    3/2\\1\\0\\0
  \end{array}
        \Rs, 
\quad
  \Delta b_{ij} = \Ls
  \begin{array}{cccc}
    9/4 & 9/2 & 0 & 0 \\
    3/2 & 7   & 0 & 0 \\
    0   & 0   & 0 & 0 \\
    0   & 0   & 0 & 0   
  \end{array}
           \Rs 
&&
\mbox{for}\ L_{ R}+\bar L_{ R}, 
\\
  \Delta b_i = \Ls
  \begin{array}{c}
    1\\0\\0\\1
  \end{array}
        \Rs, 
\quad
  \Delta b_{ij} = \Ls
  \begin{array}{cccc}
    2/3 & 0 & 0 & 16/3 \\
    0   & 0 & 0 & 0   \\
    0   & 0 & 0 & 0   \\
    2/3 & 0 & 0 & 34/3
  \end{array}
           \Rs 
&&
\mbox{for}\ H_{ T}+\bar H_{ T}, 
\\
  \Delta b_i = \Ls
  \begin{array}{c}
    0\\1\\1\\0
  \end{array}
        \Rs, 
\quad
  \Delta b_{ij} = \Ls
  \begin{array}{cccc}
    0 & 0 & 0 & 0 \\
    0 & 7 & 3 & 0 \\
    0 & 3 & 7 & 0 \\
    0 & 0 & 0 & 0  
  \end{array}
           \Rs 
&&
\mbox{for}\ H_{ D}, 
\\
  \Delta b_i = \Ls
  \begin{array}{c}
    0\\0\\0\\3
  \end{array}
        \Rs, 
\quad
  \Delta b_{ij} = \Ls
  \begin{array}{cccc}
    0 & 0 & 0 & 0 \\
    0 & 0 & 0 & 0 \\
    0 & 0 & 0 & 0 \\
    0 & 0 & 0 & 54  
  \end{array}
           \Rs 
&&
\mbox{for}\ G, 
\\
  \Delta b_i = \Ls
  \begin{array}{c}
    0\\0\\2\\0
  \end{array}
        \Rs, 
\quad
  \Delta b_{ij} = \Ls
  \begin{array}{cccc}
    0 & 0 & 0  & 0 \\
    0 & 0 & 0  & 0 \\
    0 & 0 & 24 & 0 \\
    0 & 0 & 0  & 0  
  \end{array}
           \Rs 
&&
\mbox{for}\ W_{ L}, 
\\
  \Delta b_i = \Ls
  \begin{array}{c}
    0\\2\\0\\0
  \end{array}
        \Rs, 
\quad
  \Delta b_{ij} = \Ls
  \begin{array}{cccc}
    0 & 0  & 0 & 0 \\
    0 & 24 & 0 & 0 \\
    0 & 0  & 0 & 0 \\
    0 & 0  & 0 & 0  
  \end{array}
           \Rs 
&&
\mbox{for}\ W_{ R}, 
\\
  \Delta b_i = \Ls
  \begin{array}{c}
    4\\0\\0\\1
  \end{array}
        \Rs, 
\quad
  \Delta b_{ij} = \Ls
  \begin{array}{cccc}
    32/3 & 0 & 0 & 64/3 \\
    0    & 0 & 0 & 0    \\
    0    & 0 & 0 & 0    \\
    8/3  & 0 & 0 & 34/3
  \end{array}
           \Rs 
&&
\mbox{for}\ U+\bar U, 
\\
  \Delta b_i = \Ls
  \begin{array}{c}
    4\\6\\6\\4
  \end{array}
        \Rs, 
\quad
  \Delta b_{ij} = \Ls
  \begin{array}{cccc}
    8/3 & 12 & 12 & 64/3 \\
    4   & 42 & 18 & 32   \\
    4   & 18 & 42 & 32   \\
    8/3 & 12 & 12 & 136/3
  \end{array}
           \Rs 
&&
\mbox{for}\ X_Q+\bar X_Q.
\eeqn

\end{document}